\documentclass[10pt,bezier]{article}
\usepackage{amsmath,amssymb,amsfonts, euscript}
\newcommand{\real}{I\!\!R}
\newcommand{\A}{{\bf A}}
\newcommand{\F}{{\bf F}}
\newcommand{\X}{\frak{X}}
\newcommand{\n}{\hat \nabla}
\newtheorem{thm}{Theorem}[section]

\newtheorem{example}[thm]{Example}
\newtheorem{defn}[thm]{Definition}

\newtheorem{cons}[thm]{Consequence}
\newtheorem{prop}[thm]{Proposition}

\begin{document}
\noindent {\bf\Large Affine metrics and algebroid structures: Application to general relativity and unification}\\

\noindent \\

\noindent {\bf N. Elyasi}, {\bf  N. boroojerdian}\footnote{\noindent N. Elyasi,  N. boroojerdian\\
 Dep. of Math. \& Comp., Amirkabir
University of Technology, Tehran, Iran \\
  E-mail:elyasi@yahoo.com, and broojerd@aut.ac.ir}\\

\noindent \\

\noindent {\bf  Abstract}$\quad$ Affine metrics and its associated algebroid bundle are developed.
 Theses structures are applied to the general relativity and provide an structure for unification
 of gravity and electromagnetism. The final result is a field equation on the associated algebroid bundle that
 is similar to Einstein field equation but contain Einstein field equation and Maxwell equations simultaneously
 and contain a new equation that may have new results. \\


\section{Introduction}
Different phenomena can be explained by one theory. This is the main stream of thought in physics. After the invention
of GR by Einstein in 1914, the most important question was " Is it possible to combine gravity and electromagnetism
into a unique theory?" This question is also important today, because any unification theory can through light on the
nature of the forces.

Einstein hoped that a good unification theory can solve mysteries of quantum effects and elementary particles too, and
spent decades of his life on this project. Of course, he was unsuccessful and it does not seem  that the mathematical
framework of classical physic be suitable for explanation of quantum effects and elementary particles. To comprehend
quantum phenomena, we need to change our view points drastically. So, a unification theory of gravity and
electromagnetism in the level of classical physic can not be considered as a true theory and at most we can expect
that it be a good approximation in the level of classical physic of the true theory.

The main idea in the unification of gravity and electromagnetism or geometrization of electromagnetism is to provide
new geometrical structures such that simultaneously contain both gravitation and electromagnetism naturally. It has
been done many attempts to produce convenient geometrical structures. Most of them consider GR as a base structure and
enrich it with additional structures or additional degrees of freedom.

Early attempts in unification of fields have been done by Weyl(considering conformal structures
and introducing a new gauge transformation, 1918), Kaluza (adding additional dimension to
space-time, 1919), Eddington(considering connection as the central concept and decomposing its
Ricci tensor to symmetric and anti-symmetric parts,  1921), Schouten(considering connections with
nonzero torsion, 1921),Klein(interpreting fifth dimension of Kaluza theory as a relation to
quantum concepts, 1926), Infeld(considering asymmetric metric that its symmetric part represent
gravity and its anti-symmetric part represent electromagnetism field, 1928), Einstein and
Mayer(5-vector formalism and considering vector bundles and connections on vector bundles,
1931)[3].

 The main idea of this paper is to to introduce affine metrics and its associated algebroid bundle on a space-time.
connections and curvature of theses structures naturally contain elements of gravity and
electromagnetism and a field equation can relate these concepts properly.

\section{Algebraic preliminary}
Let $V$ and $W$ be vector spaces, then a function $S:V\longrightarrow W$ is called an affine function iff there exists
a linear function $T:V\longrightarrow W$ such that:
\[
  \forall u,v\in V\hspace{2mm} S(u+v)=S(u)+T(v)
\]
 $T$ is unique and is called the linear part of $S$. All affine function $S:V\longrightarrow W$ have the form $S(u)=T(u)+a$ in
 which $T$ is linear and $a\in W$. In this section $V$  is a fixed vector space.

\subsection{2-affine functions and affine inner products}
\begin{defn}
 A function $S:V\times V\longrightarrow \real$ is called 2-affine iff it is affine in each argument. And $S$ is
  called affine-linear iff it is affine in first variable an linear in second
  variable. In a similar way $S$ is called linear-affine iff it is linear in first and affine in second variable.
\end{defn}
\begin{defn}
A 2-affine function $S:V\times V\longrightarrow \real$ is called symmetric iff for all $a,b\in E:\ S(a,b)=S(b,a)$
\end{defn}
If $S:V\times V\longrightarrow \real $ be a symmetric 2-affine function then there exist a unique linear-affine
function $T_1$
 and a unique symmetric bilinear function $T$ such that for all $a,b,u,v\in V$:
\[
 S(a+u,b+v)=S(a,b)+T_1(u,b)+T_1(v,a)+T(u,v)
\]
  $T_1$  and $T$ are called respectively linear-affine, and bilinear parts of $S$.

\begin{defn}
A 2-affine function \mbox{$S:V\times V \longrightarrow \real $} is called an affine inner product on $V$, iff $S$ be
symmetric and its bilinear part be an inner product on V.
\end{defn}
Every ordinary inner product on  $V$, is also an affine inner product.

\noindent {\bf Notation:} Suppose $a,b,u,v\in V$. Usually, we show an affine inner product by $(a,b)$ and its
linear-affine  parts  by $<u,b)$ and  its bilinear part by $<u,v>$. For simplicity in writing, we also use an
affine-linear function denoted by $(b, u>$ that is equal to $<u,b)$. So,
\[
(a+u,b+v)=(a,b)+(a,v>+<u,b)+<u,v>
\]
Let $V$ be a vector space and $(.,.)$ be an affine inner product on $V$, then for a unique vector ${\bf z}\in V$
and a unique scalar $\lambda\in \real$ we have
\[
(u,v)=\lambda+<u-{\bf z},v-{\bf z}>
\]
It is sufficient to set ${\bf z}$ be the vector that for all $v\in V$, $<{\bf z},v>=-(0,v>$ and $\lambda=(0,0)-<{\bf
z},{\bf z}>$. Conversely for any ordinary inner product on $V$ and vector ${\bf z}\in V$ and scalar $\lambda\in\real$,
by the above formula we can define an affine inner product on $V$ and all affine inner products on $V$ are obtained in
this way.

\subsection{associated inner product space to affine metrics}
Let $(,)$ be an affine inner product on $V$. Set $\widehat V$ be the space of real valued affine map on $V$. $\widehat
V$ is a vector space whose dimension is one plus dimension of $V$. For all $x\in V$ set $\hat x:V\longrightarrow\real$
be the affine map $\hat x(y)=(x,y)$. The map $x\mapsto \hat x$ is affine and imbed $V$ into $\widehat V$ as an affine
subspace.

 For all $x\in V$ set $\bar x:V\longrightarrow\real$ be the affine map $\bar x(y)=<x,y)$.
The map $x\mapsto \bar x$ is linear and imbed $V$ into $\widehat V$ as a vector subspace. Denote the set of all $\bar
x$ by $\bar V$. The space of real valued constant function on $V$ is a one dimensional subspace of $\widehat V$ and is
complementary to $\bar V$. So, we find a natural projection $\rho :\widehat V\longrightarrow V$ whose kernel is
constant functions and its restriction to $\bar V$ is $\bar x\mapsto x$.

If $(x,y)=\lambda +<x-{\bf z},y-{\bf z}>$ and $\lambda\neq 0$, then there exist a unique inner
product on $\widehat V$ such that for all $x,y\in V$ we have $<\hat x,\hat y>=(x,y)$. By
manipulating this property we can find the right definition of this inner product. $\hat{\bf z}$
is the constant function $\hat{\bf z}(x)=\lambda$ and must be orthogonal to $\bar V$. Every
element of $\widehat V$ is uniquely written in the form $\bar x+\mu\hat{\bf z}$, and we must define
\[
<\bar x+\mu_1\hat{\bf z},\bar y+\mu_2\hat{\bf z}>=<x,y>+\lambda\mu_1\mu_2
\]
Note that for all $x,y\in V$ we have $\hat x+\bar y=\widehat{x+y}$, so $\hat x=\overline{x-z}+\hat
{\bf z}$.

\section{Affine semi-riemannian manifolds and its associated algebroid}
 In this section, $M$ is a fixed smooth manifold and all functions are smooth.
\begin{defn}
If for every $p\in M$, we choose on every $T_pM$ an affine inner product smoothly, then we call it an affine
metric on $M$ and $M$ is called an affine semi-riemannian manifold.
\end{defn}
Every semi-riemannian manifold is also an affine semi-riemannian manifold. The bilinear part of an affine metric
on $M$ is a semi-Riemannian metric on $M$ and it is called the associated semi-Riemannian metric.
\begin{example}
If $<.,.>$ be a semi-Riemannian metric on $M$, and ${\bf A}\in\frak{X}M$ and \mbox{$\phi\in C^\infty(M)$}, then the
following formula defines an affine metric on $M$.
\[\forall X,Y\in \frak{X}M\hspace{5mm}(X,Y)=\phi\ +<X-{\bf A},Y-{\bf A}>\]
\end{example}
Every affine metric on $M$ can be written as above. If ${\bf 0}$ be the zero vector field, it is sufficient to set
${\bf A}$ be the vector field which for all $X\in\frak{X}M$,  \mbox{$<{\bf A},X>=-({\bf 0},X>$} in which $<.,.>$ is the
bilinear part of the affine metric and set $\phi=({\bf 0},{\bf 0})-<{\bf A},{\bf A}>$. In this section $M$ is an
affine semi-riemannian manifold and $X,Y\in\X M$ and its affine metric is as follows:
\[(X,Y)=1\ +<X-{\bf A},Y-{\bf A}>\]
Let $\widehat {TM}$ be the vector bundle $\bigcup_{p\in M}\widehat{T_pM}$. For every $X\in\frak XM$
let $\hat X$ and $\bar X$ be sections of $\widehat {TM}$ such that $(\hat X)_p=\widehat{X_p}\ ,\
(\bar X)_p=\overline{X_p}$. $\hat {\bf A}$ is the constant function $\hat{\bf A}(u)=1$ and for
simplicity we denote it by $\xi$. Let $\overline {TM}$ be the vector bundle $\bigcup_{p\in
M}\overline{T_pM}$ that is a subvector bundle of $\widehat {TM}$. $\overline {TM}$ is
complementary to line subbundle generated by ${\bf \xi}$. Every section of $\widehat {TM}$
uniquely written in the form $\bar X+f\xi$ for some $X\in\frak XM$ and $f\in C^\infty (M)$.
$\widehat {TM}$ is a semi-riemannian vector bundle by the induced inner product:
\[ X,Y\in\frak XM,f,g\in C^\infty (M)\quad <\bar X+f\xi,\bar Y+g\xi>=<X,Y>+fg\]
$\widehat {TM}$ has a natural algebroid structure over $TM$. The anchor map is
\[\begin{array}{rccc}
\rho :&\widehat{TM}&\longrightarrow& TM\\
 &\bar X+f\xi&\longmapsto &X
\end{array}\]
In the definition of Lie bracket on $\widehat {TM}$ the vector field $\A$ make a crucial role. $\nabla\A$ is a 1-1
tensor on $M$ and its anti symmetric part is denoted by $\F$. In fact:
\[ <\F (X),Y>={1\over 2}(<\nabla_X\A,Y>-<X,\nabla_Y\A>) \]
$2\F$ is equivalent to the exterior derivation of the 1-form equivalent to $\A$. Lie bracket on  $\widehat {TM}$ is
defined as follows:
\[
 [\bar X,\bar Y]=\overline {[X,Y]}+2<\F (X),Y>\xi \quad , \quad [\bar X,\xi ]={\bf 0}
\]
Jacobi identity is hold because $\F$ is equivalent to a closed form. Now,  $\widehat {TM}$ is a semi-riemannian
algebroid over $TM$ and has a unique Levi-civita connection $\n$ which can be computed by the following relation[5].
For all $ U,V,W\in\Gamma (\widehat{TM})$:
\[\begin{array}{rl}
 2<\n_UV,W> =& \rho (U)<V,W>+\rho (V)<W,U>-\rho (W)<U,V>\\
 & +<[U,V],W>-<[V,W],U>+<[W,U],V>
\end{array}\]
\begin{prop}
Levi-civita connection of the algebroid $\widehat {TM}$ satisfies the following relations.
\end{prop}
\[\begin{array}{rcl}
\n_\xi\xi&=&{\bf 0} \\    \n_{\bar X}\xi&=&\n_\xi\bar X\ =\ -\overline{\F (X)} \\
\n_{\bar X}\bar Y&=&\overline{\nabla_XY}+<\F (X),Y>\xi
\end{array}\]
{\bf proof:} Straightforward computations show these results. $\blacksquare$
\begin{defn}
A path $\hat\alpha :I\longrightarrow\widehat{TM}$ is called a geodesic of $\n$ iff for some path $\alpha
:I\longrightarrow M$ we have $\rho (\hat\alpha)=\alpha'$ and $\n_{\hat\alpha}\hat\alpha =0$.
\end{defn}
\begin{prop}
A path $\hat\alpha :I\longrightarrow\widehat{TM}$ is a geodesic of $\n$ iff for some path $\alpha :I\longrightarrow M$
and scalar $\lambda$ we have $\hat\alpha(t)=\overline{\alpha'(t)}+\lambda\xi$ and
$\nabla_{\alpha'}\alpha'=2\lambda\F(\alpha')$.
\end{prop}
{\bf Proof:} Since $\rho (\hat\alpha)=\alpha'$, for some function $f:I\longrightarrow\real$ we have $ \hat\alpha(t)=
\overline{\alpha'(t)}+f(t)\xi$. To compute easily, assume $X$ be a local vector field on $M$ and $g$ a local function
on $M$ such that $X_{\alpha(t)}=\alpha'(t)$ and $g(\alpha(t))=f(t)$. Consequently, $\hat \alpha(t)=(\bar
X+g\xi)_{\alpha(t)}$. Since $\n_{\hat\alpha}\hat\alpha =0$, we have:
\[\begin{array}{rl}
0&=\n_{\hat\alpha}\hat\alpha =(\n_{\bar X+g\xi}\bar X+g\xi)_{\alpha(t)}=(\n_{\bar X}\bar X+g\n_\xi\bar X+\n_{\bar X}
g\xi+g\n_\xi g\xi)_{\alpha(t)}\\
 & =(\overline{\nabla_XX}-2g\overline{\F(X)}+X(g)\xi)_{\alpha(t)}\\
 &\Rightarrow\quad \nabla_{\alpha'(t)}\alpha'(t)=2f(t)\F(\alpha'(t))\ ,\ \alpha'(t)(g)=0
\end{array}\]
second equation means $g(\alpha(t))'=f'(t)=0$, so $f$ is constant and for some scalar $\lambda,\ f(t)=\lambda$.
Consequently $\hat\alpha(t)=\overline{\alpha'(t)}+\lambda\xi$ and
$\nabla_{\alpha'}\alpha'=2\lambda\F(\alpha').\blacksquare$
\begin{prop}
The curvature tensor of $\n$ denoted by $\hat R$, and it satisfies the following relations.
\end{prop}
\[\begin{array}{rcl}
\hat R(\bar X,\xi)(\xi)&=&-\overline{\F(\F(X))} \\
\hat R(\bar X,\xi)(\bar Z)&=&-\overline{(\nabla_X\F)(Z)}-<\F (X),\F (Z)>\xi \\
\hat R(\bar X,\bar Y)(\xi)&=&-\overline{(\nabla_X\F)(Y)}+\overline{(\nabla_Y\F)(X)} \\
\hat R(\bar X,\bar Y)(\bar Z) &=& \overline{R(X,Y)(Z)}+<Z,\F (X)>\overline{\F (Y)}-<Z,\F (Y)>\overline{\F (X)}\\
 & & +2<\F (X),Y>\overline{\F (Z)}+<Z,\overline{(\nabla_X\F)(Y)}-\overline{(\nabla_Y\F)(X)}>\xi
\end{array}\]
{\bf proof:} Straightforward computations show these results. $\blacksquare$
\begin{prop}
The Ricci curvature tensor of $\n$ denoted by $\widehat {Ric}$, as a 1-1 tensor satisfies the following relations.
\end{prop}
\[\begin{array}{rcl}
\widehat {Ric}(\xi)&=&\overline{div(\F)}-tr(\F\circ\F)\xi \\
\widehat {Ric}(\bar X)&=&\overline{Ric(X)}+2\overline{\F (\F (X))}+<div(\F),X>\xi
\end{array}\]
{\bf proof:} Let $X_1,\cdots,X_n$ be an orthonormal local base for $M$. So, $\overline{X_1},\cdots ,\overline{X_n},\xi$
is an orthonormal local base for $\widehat {TM}$. Set $\epsilon_i=<X_i,X_i>=\pm 1$, note that for any vector $Y\in\X M$
we have $Y=\sum_i\epsilon_i<X_i,Y>X_i$.
\[\begin{array}{rcl}
\widehat {Ric}(\xi) &=& \sum_i\epsilon_i\hat R(\xi,\overline{X_i})(\overline{X_i})+\hat R(\xi,\xi)(\xi)\\
 & =&\sum_i\epsilon_i \overline{(\nabla_{X_i}\F)(X_i)}+\sum_i\epsilon_i <\F (X_i),\F (X_i)>\xi\\
 &=&\overline{\sum_i\epsilon_i (\nabla_{X_i}\F)(X_i)}-\sum_i\epsilon_i<\F (\F (X_i)),X_i>\xi \\
 &=& \overline{div(\F)}-tr(\F\circ\F)\xi
\end{array}\]
\[\begin{array}{rcl}
\widehat {Ric}(\bar X) &=& \sum_i\epsilon_i\hat R(\bar X,\overline{X_i})(\overline{X_i})+\hat R(\bar X,\xi)(\xi)\\
 & =&\sum_i\epsilon_i\big( \overline{R(X,X_i)(X_i)}+<X_i,\F (X)>\overline{\F (X_i)}-<X_i,\F (X_i)>\overline{\F (X)}\\
 & & +2<\F (X),X_i>\overline{\F (X_i)}+<X_i,\overline{(\nabla_X\F)(X_i)}-\overline{(\nabla_{X_i}\F)(X)}>\xi \big)
 -\overline{\F (\F (X))}\\
 &=& \overline{Ric(X)}+3\overline{\F(\sum_i\epsilon_i<X_i,\F
 (X)>X_i)}-\sum_i\epsilon_i<X_i,\overline{(\nabla_{X_i}\F)(X)}>\xi -\overline{\F (\F (X))}\\
 &=&\overline{Ric(X)}+3\overline{\F(\F(X))}+\sum_i\epsilon_i<X,\overline{(\nabla_{X_i}\F)(X_i)}>\xi -\overline{\F (\F
 (X))}\\
&=&\overline{Ric(X)}+2\overline{\F (\F (X))}+<div(\F),X>\xi\quad \blacksquare.
\end{array}\]
Denote the scalar curvature of $\widehat {TM}$ by $\hat R$. An easy computation shows that:
\[\hat R=R+tr(\F\circ\F)\]

\section{Application to general relativity }
In the rest of the paper, suppose $M$ is a four manifold and $X,Y\in\frak{X}M$.
\begin{defn}
An affine metric $(,)$ on $M$ is called affine-Lorentzian iff its bilinear part be a Lorentzian metric and for some
$\A\in\X M$:
\[(X,Y)=1\ +<X-{\bf A},Y-{\bf A}>\]
\end{defn}
An affine Lorentzian metric is determined by a Lorentzian linear metric $<.,.>$ and a vector field ${\bf
A}\in\frak{X}M$. It seems plausible to interpret $<.,.>$ as a potential for gravity and ${\bf A}$ as a potential for
electromagnetism.

These interpretations can be rational, if we can interpret projection of the geodesics  of $\widehat{TM}$ to $M$
 as world-lines of charged particles. By proposition (3.5) geodesics of $\widehat{TM}$ determine paths $\alpha$ on $M$
 that satisfy equation $\nabla_{\alpha'}\alpha'=2\lambda\F(\alpha')$. $2\F$ is the electromagnetism  tensor with potential
 $\A$ and this is exactly the equation of a charged particle that its ration of charge to mass is $\lambda$ and move under
  influence of electromagnetism tensor $2\F$.

\subsection{Field equation}  We need a field equation that naturally contains Einstein and maxwell equations simultaneously.
To have a sound formulation, we use a system of measurement in which $c=1$, $G=1$, and $\epsilon_0={\textstyle
1\over\textstyle 16\pi}$. In this system, Einstein field equation and maxwell equation are written as following[1]:
\begin{eqnarray*}
Ric-{1\over 2}R{\bf g}&=&8\pi ({\bf T}^{mas}+{\bf T}^{elec})\\
div(2{\bf F})&=&16\pi {\bf J}
\end{eqnarray*}
 So, $div(\F)=8\pi {\bf J}$. In this system, the momentum-energy tensor of the electromagnetism field is as following:
 \[
 {\bf T}^{elec}_{ij}={\textstyle 1\over\textstyle  16\pi}(2\F_{im}2\F^m_j-{\textstyle 1\over\textstyle 4}
 {\bf g}_{ij}2\F_{mn}2\F^{mn})={\textstyle 1\over\textstyle  4\pi}(\F_{im}\F^m_j-{\textstyle 1\over\textstyle 4}
 {\bf g}_{ij}\F_{mn}\F^{mn})
 \]
In fact ${\bf T}^{elec}={\textstyle -1\over\textstyle  4\pi}(\F\circ\F-{\textstyle 1\over\textstyle
4}tr(\F\circ\F){\bf g})$.

 Einstein field equation can be rewritten as following:
\[\begin{array}{rl}
 &Ric-{1\over 2}R{\bf g}=8\pi ({\bf T}^{mas}-{\textstyle 1\over\textstyle  4\pi}(\F\circ\F-{\textstyle 1\over\textstyle
 4}tr(\F\circ\F){\bf g}))\\
\Rightarrow & Ric-{1\over 2}R{\bf g}+2(\F\circ\F-{\textstyle 1\over\textstyle  4}tr(\F\circ\F){\bf g})=8\pi {\bf T}^{mas}\\
\Rightarrow & Ric+2\F\circ\F-{1\over 2}(R+tr(\F\circ\F)){\bf g}=8\pi {\bf T}^{mas}
\end{array}\]
If we construct Einstein tensor $\hat G=\hat{Ric}-{1\over 2}\hat R{\bf\hat g}$ in the algebroid $\widehat{TM}$ we can
see that the left side of Einstein field equation is exactly the restriction of $\hat G$ to $\overline{TM}$. It seems
that we can construct a suitable field equation in $\widehat{TM}$ such that contains simultaneously Einstein and
maxwell field equations. We need a proper 5-mass momentum-energy tensor $\hat T$ such that the equation:
\[
\widehat{Ric}-{1\over 2}\hat R{\bf\hat g}=8\pi \hat T
\]

represent both Einstein and maxwell  equations. Left side of this equation is a geometrical object and contains both
gravity and electromagnetism as geometrical objects. Right side of this equation must contain information about matter
that determine geometry of space. If $T$ be ordinary mass momentum-energy tensor we must have:
\[
\hat T(\bar X,\bar Y)=T(X,Y)
\]
Since $\hat G(\bar X ,\xi)=<\widehat{Ric}(\bar X),\xi>-{1\over 2}\hat R<\bar X ,\xi>=<div(\F),X>$, we must have
$<div(\F),X>=8\pi \hat T(\bar X,\xi)$. But by maxwell equations we have $div(\F)=8\pi {\bf J}$, so we must define:
\[ \hat T(\bar X,\xi)= \hat T(\xi , \bar X)=<X,{\bf J}> \]
Definition of $\hat T(\xi,\xi)$ is not straightforward. We need some clue to find right definition. Put $H=\hat
T(\xi,\xi)$, so we can consider $({\bf J},H)$ as 5-current. Let $\eta$ be the charge density and $\rho$ be the mass
density and ${\bf U}$ be 4-velocity. If $({\bf U}, U_5)$ be 5-velocity, 5-current has the form $({\bf J},H)=\eta({\bf
U},U_5)$ and $\hat T$ has the form $\rho U_iU_j$. So we must have $\eta U_5=\rho U_5U_5$, and consequently
$U_5={{\textstyle \eta}\over {\textstyle \rho}}$. So, fifth component of 5-velocity must be the ratio of charge to
mass. Some other clues support this reason. In computing geodesics of $\widehat {TM}$ we find that $\hat\alpha$ that
is 5-velocity of the geodesic $\alpha$ has the form $\hat\alpha(t)=\overline{\alpha'(t)}+\lambda\xi$. Fifth component
of this 5-velocity is $\lambda$ and it was the ratio of charge to mass.

By this result we can conclude $\hat T(\xi,\xi)=H=\eta U_5={{\textstyle \eta^2}\over{\textstyle\rho}}$. Now, we ready
to write field equation in the algebroid $\widehat {TM}$:
\[
\widehat{Ric}-{1\over 2}\hat R{\bf\hat g}=8\pi \hat T
\]
By spiliting $\widehat {TM}$ to subbundle $\overline {TM}$ and line bundle generated by $\xi$ the above equation can
be written in the following block form.
\[
\left( \begin{array}{c|c}
\begin{array}{c}  Ric+2\F\circ\F\\  -{1\over 2}(R+tr(\F\circ\F)){\bf g}  \end{array} & div(\F) \\ \hline div(\F) &
 -{1\over 2}(R+3tr(\F\circ\F))
\end{array}\right) =8\pi
\left( \begin{array}{c|c} T^{mass} & {\bf J} \\ \hline {\bf J} & {{\textstyle \eta^2}\over{\textstyle\rho}}
\end{array}\right)
\]
This equation produce three other equations, that two of them are Einstein and maxwell field equations and third
equation is new:
\[ R+3tr(\F\circ\F)=-16\pi {{\textstyle \eta^2}\over{\textstyle\rho}} \]

\section{conclusion}
algebroid structures and affine metrics provide a mathematical framework for unification of gravity and
electromagnetism. This theory is very similar to Kaluza theory except that it need not an extra dimension in base
manifold. Instead, affine metrics naturally produce an extra dimension in tangent space. So, many  redundancy  of
Kaluza theory disappear but one new equation appears that shows an intimate relation between mass and charge densities.


\begin{thebibliography}{12}
\bibitem{} Sachs, R.K. and  Wu, H.:  General Relativity for Mathematicians, Springer-verlag, New York, 1977.
\bibitem{} W. A. Poor.: Differential Geometric Structures, McGraw-Hill, 1981.
\bibitem{}Hubert F. M. Goenner.: On the History of Unified Field Theories, Max Planck
 Institute for Gravitational Physics Albert Einstein Institute, (2004).
\bibitem{} Myroon, W. Evans: Generally covariant Unified Field Theory, abramis (2005)
\bibitem{}Mohamed Boucetta: Riemannian Geometry of Lie algebroids, arxiv:0806.3522v2(2008)
\bibitem{}K.Grabowska, J.Grabowski, P.Urbanski: Geometrical Mechanics on algebroids, int. J. Geom. Method. Phys. 3(2006)
\bibitem{}J. M. overdain: Kaluza-Klien Gravity, arxive:gr-qc/9805018v1(1998)
\bibitem{}Paul S Wesson: space time matter: Modern Kaluza-Klien theory, world scientific
\end{thebibliography}
 \end{document}